\documentclass[12pt,a4paper]{article}
\usepackage{amsthm}
\usepackage{amsmath}
\usepackage{natbib}
\usepackage[colorlinks,citecolor=blue,urlcolor=blue,filecolor=blue,backref=page]{hyperref}
\usepackage{graphicx}
\usepackage{xcolor}
\usepackage{algorithm}
\usepackage{adjustbox}
\usepackage[noend]{algpseudocode}
\usepackage{natbib}
\usepackage{bbm, dsfont}
\usepackage{amsfonts}
\usepackage{siunitx}
\usepackage[paperheight=13in,paperwidth=9.5in]{geometry}

\newcommand{\be}{\begin{equation}}
\newcommand{\ee}{\end{equation}}
\newtheorem{remark}{\textit{Remark}}

\begin{document}

\title{\textbf{\Large On the estimation of population size from a post-stratified two sample capture-recapture data under dependence}}

\date{}
\maketitle
\author{
\begin{center}
\vskip -1cm

Kiranmoy Chatterjee \\
Interdisciplinary Statistical Research Unit, Indian Statistical Institute\\
E-mail: kiranmoy07@gmail.com\\

Prajamitra Bhuyan\\
Department of Mathematics, Imperial College London\\
E-mail: bhuyan.prajamitra@gmail.com
\end{center}
}

\begin{abstract}
Population size estimation based on two sample capture-recapture type experiment is an interesting problem in various fields including epidemiology, pubic health, population studies, etc. The Lincoln-Petersen estimate is popularly used under the assumption that capture and recapture status of each individual is independent. However, in many real life scenarios, there is an inherent dependency between capture and recapture attempts which is not well-studied in the literature of the dual system or two sample capture-recapture method. In this article, we propose a novel model that successfully incorporates the possible causal dependency and provide corresponding estimation methodologies for the associated model parameters based on post-stratified two sample capture-recapture data. The superiority of the performance of the proposed model over the existing competitors is established through an extensive simulation study. The method is illustrated through analysis of some real data sets.
\end{abstract}

{\bf Keywords :} 	Behavioural dependency, Bivariate Bernoulli, Disease surveillance, Method of moments, Maximum likelihood, Post-stratification.  \\

	\section{Introduction}\label{Intro}
Estimation of the size of a population is an interesting problem in different disciplines of epidemiological, medical, social and demographic studies. In order to formulate policies for public heath related issues, federal agencies are generally interested to know the actual size of a diseased population (e.g. Encephalitis patients) or vital events (e.g. child mortality) in a specified region. Any attempt to count all the individuals belonging to a population of interest is always subject to error and the degree of error depends on many factors, such as, population size, individual's capture probability, etc. In this context, two sources of information have extensive use for human population as more than two sources are hardly found in demographic study due to various practical constraints such as survey cost, human mobility, etc. \citep{Chatterjee16c}. In order to draw inference from two capture attempts, one needs to combine the data obtained from the two surveys and determine how many people are included in both the lists and how many are included exactly in one of the lists. Therefore, an incomplete $2\times2$ cross-classified data structure is obtained and it is known as dual-record system (DRS). This data structure is similar to the two sample capture-recapture data \citep{Wolter86, Chatterjee16a}. In DRS, counts for the three cells are available, however the last cell count remained unknown which makes the true population size, say $N$, unknown. The primary goal is to estimate the missing cell count, or equivalently $N$, from the available data. This is somewhat close to the capture-recapture experiment, widely practiced in wild-life studies, with only one recapture attempt. Often, survey mechanism allows post-stratification of the entire population into mutually exclusive and exhaustive sub-populations based on demographic and social characteristics (e.g. age, sex, ethnicity, etc.), and it is also of great interest to estimate the sub-population sizes \citep{Bell93, Wolter90}.

In order to estimate $N$, a common practice is to assume causal independence between capture and recapture attempts and the resulting estimator is popularly known as Lincoln-Petersen (LP) estimator in DRS \citep{Otis78, Bohning09}. With an additional assumption of \textit{time-variation}, \citet{Otis78} proposed the model $M_t$ and the resulting estimator is same as LP estimator. \citet{Chatterjee16a} proposed an integrated likelihood estimation methodology based on the model $M_{t}$ and compared its performance with others likelihood based estimators in DRS. However, model $M_t$ (equivalently, the LP estimator) often fails due to positive dependence among the two lists, especially in the fields of public health and demography, which leads to underestimation of $N$ \citep{Hook82, Chao01a}. For example, patients with positive result from a serum test for Hepatitis A Virus (HAV) are prone to visit hospital for further treatment. Therefore, the ascertainment of the serum sample and that of the hospital sample becomes dependent. In census-undercount study, \citet{Fay88} and \citet{Bell93} observed such dependence in behavioral response among adult males but not for females in the Post Enumeration Programs conducted for evaluating the US Censuses in 1980 and 1990 respectively. In epidemiological or demographic surveillance with two sample capture-recapture experiment, positive list-dependence is often observed \citep{Chatterjee16c,Schrauder07}. Similarly, there are some populations in which negative dependence is encountered, such as children injury data collected by hospitals and police stations, drug abused population, population of patients affected with HIV or any other diseases that bear social stigma \citep{Chatterjee16c,Chatterjee18}. Recently, \citet{Yang10} have proposed an empirical Bayes estimator which performs better than LP estimator as well as some of its modified versions including Chapman's and Bailey's estimators. However, their underlying hypergeometric model does not encounter the list-dependence. In this context, model $M_{tb}$, proposed by \citet{Otis78}, exclusively includes the list-dependence in terms of behavioral response effect parameter, but this model is not estimable in DRS \citep{Chao00,Chatterjee16c}. \citet{Yang05} proposed a Markov chain approach that incorporates both long-term as well as short-term behavioral response effects in the existing models for capture-recapture experiments. However, their model is also not estimable in DRS \citep{Chatterjee15}. 

Modeling of the capture-recapture data incorporating the causal dependence assumption is an important but challenging task in DRS. \citet{Nour82} proposed an estimate of total number of vital records assuming the positive dependence between two lists in DRS of vital events registration. \citet{Wolter90} provided estimation for post-strata wise sub-population (e.g. male, female) sizes under two different models, assuming the ratio of the sub-population sizes (i.e. the sex-ratio) to be known from Demographic Analysis. In the first model, \citet{Wolter90} considered that the cross-product ratios in DRSs for male and female post-strata are same but unknown and, in the second one, causal independence is assumed for the female only. \citet{Isaki86} also worked on the same problem for 1980 Post Enumeration Program and suggested an estimate based on demographic analysis. Later, \citet{Bell93} proposed some variations of the methods suggested by \citet{Wolter90} for the estimation of the cross-product ratios for both male and female populations. However, the ratio of the sub-population sizes (e.g. sex-ratio) is calculated at the time of census for larger population (e.g. national level population). In many situations, it is not realistic to assume that the ratio remains constant over time or holds true for the sub-populations under consideration. Moreover, the availability of this ratio for the population of interest is very much limited across the various fields where the DRS type data structure is commonly used (e.g., epidemiological or disease surveillance data; \textit{See} Section \ref{realdata}).

In this article, we propose a novel model to incorporate this inherent dependency between capture and recapture attempts in DRS without the knowledge on the ratio of the sub-population sizes and provide estimation methodologies for the population size $N$ based on post-stratification under two different scenarios. Our model can also incorporate available information on the ratio of the sub-population sizes and provides better result than the existing competitor. Our work is motivated from two real datasets on public health: (\textit{i}) \textit{Encephalitis incidence} in England, 2006-2007 and (\textit{ii}) \textit{child mortality} in western Kenya, 2000-2001, where the existing methods proposed by \citet{Wolter90} and \citet{Nour82} are not applicable. Our model possesses nice interpretation, and associated estimates exhibit superiority with respect to relative bias, relative root mean squared error and coverage probability over the existing competitors available in the literature (\textit{See} Section \ref {est}, \ref{simu}). We first describe the DRS and the associated data structure in Section 2. In Section 3, we propose a Bivariate Bernoulli model under DRS. Next, in Section 4, we derive method of moments estimates and discuss maximum likelihood estimation of the model parameters. Comparison of the proposed estimators with its existing competitors is studied through extensive simulation and two illustrative data analyses in Sections 5 and 6, respectively. Finally, we end with some concluding remarks in Section 7.

\section{Dual-record System (DRS)}\label{DRS}
As discussed in Section \ref{Intro}, DRS is similar to the two sample capture-recapture sampling which is very common in estimation of the size of human population.  Let us consider a population $U$ of size \emph{N}. The individuals captured in the first list (e.g. census) are matched one-by-one with the individuals captured in the second list (e.g. Post Enumeration Program). Let $p_{j 1\cdot}$ and $p_{j\cdot1}$ denote the capture probabilities of the \emph{j}th individual in the first sample (List 1) and the second sample (List 2), respectively. Under this set-up, we consider the following assumptions:\\

($S1$) population is closed until the second sample is taken,\\

($S2$) individuals are homogeneous with respect to their capture probabilities in each of the two attempts.\\
\\
Assumption ($S2$) ensures that $p_{j 1\cdot}=p_{1\cdot}$ in List 1 and $p_{j \cdot1}=p_{\cdot1}$ in List 2 for \emph{j} = 1, 2, $\cdots$, \emph{N}. The data structure, presented in Table $\ref{tab:1}$, is popularly known as the Dual-record system or shortly, DRS. The number of untapped individuals in both the surveys, denoted as $x_{00}$, is unknown which makes the total population size \emph{N} unknown. The probabilities attached to all the cells are also provided in Table $\ref{tab:1}$ and these notation will be followed throughout this paper. As discussed before, casual independence is assumed between capture and recapture attempts, which is formally written as$\colon$\\
$(S3)$ inclusion of each and every individual, belonging to $U$, in the List 2 is \textit{causally independent} to its inclusion in the List 1 (i.e. $p_{11}=p_{1\cdot}p_{\cdot1}$).\\

\begin{table}[ht]
	\centering
	\caption{Dual-record-System (DRS): $2\times2$ data structure with cell probabilities mentioned in [ ] and $p_{\cdot\cdot}$=1}
	\begin{tabular}{lccc}
		&\multicolumn{3}{c}{List 2} \\
		\cline{2-4}
		List 1 & In & out & Total\\
		\hline \hline
		In & $x_{11}[p_{11}]$ & $x_{10}[p_{10}]$ & $x_{1\cdot}[p_{1\cdot}]$\\
		Out& $x_{01}[p_{01}]$ & $x_{00}[p_{00}]$ & $x_{0\cdot}[p_{0\cdot}]$\\ 
		\hline
		Total& $x_{\cdot1}[p_{\cdot1}]$ & $x_{\cdot0}[p_{\cdot0}]$ & $x_{\cdot\cdot}=N[p_{\cdot\cdot}]$\\
		\hline
	\end{tabular}
	\label{tab:1}
\end{table}
Now assuming $(S3)$, estimate of $N$ is found as
\begin{eqnarray}\label{LP}
\hat{N}^{(LP)}=\frac{x_{1\cdot1}x_{\cdot1}}{x_{11}}, 
\end{eqnarray}	
which is popularly known as the Lincoln-Petersen (LP) estimator. This estimator is identical with the conditional likelihood estimator of $N$ from the model $M_{t}$ \citep{Wolter86} and it is traditionally used in several studies including public health, economics, demography \citep{Bohning09}. However, this model is seriously criticized due to its underlying causal independence assumption $(S3)$ in the context of human populations \citep{ChandraSekar49,Chao01a}. In many situations, failure in capturing one individual in both the attempts may be due to some common causes, and that leads to a positive association between the two lists. In some other cases, individuals may be less keen to be enlisted in List 2 which results in a negative association between the lists. These phenomena are broadly known as behavioral response variation (\textit{See} \citet{Wolter86} for more details). 

In the context of demographic studies, \citet{Nour82} considered possible positive association between the two lists in DRS. Assuming both the marginal list capture probabilities (i.e. $p_{1\cdot}$ and $p_{\cdot1}$) are greater than 0.5, \citet{Nour82} derived the estimate of $N$ as  
\begin{eqnarray}\label{Nour}
\hat{N}^{(Nour)}=x_{0}+\frac{2x_{11}x_{10}x_{01}}{x_{11}^2+x_{10}x_{01}},  
\end{eqnarray}	
where $x_{0}=x_{11}+x_{10}+x_{01}.$

As mentioned in the previous section, \citet{Wolter90} considered post-stratification of the entire population $U$ into two mutually exclusive and exhaustive sub-populations, say $U_A$ and $U_B$, (e.g., male and female) of sizes $N_A$ and $N_B$ such that $N_A+N_B=N$. Therefore, the observed data, as presented in Table \ref{DRS}, are divided into ($x_{11A},x_{10A},x_{01A}$) and ($x_{11B},x_{10B},x_{01B}$) for the two sub-populations $U_A$ and $U_B$, respectively. Based on the above datasets, \citet{Wolter90} proposed two models where one common assumption is that the ratio of the sub-population sizes $r=N_A/N_B$, is known. In the first model, say Wolter-1, the cross-product ratios for $U_A$ (say, $\theta_A=\frac{p_{11A}p_{00A}}{p_{10A}p_{01A}}$) and $U_B$ (say, $\theta_B=\frac{p_{11B}p_{00B}}{p_{10B}p_{01B}}$), are assumed to be same but unknown, i.e., $\theta_A=\theta_B$. The estimates of the sub-population sizes from Wolter-1 are given by
\begin{eqnarray}\label{WolterM1}
\hat{N}_{B}^{(W1)}=max\left(\frac{Qx_{0B}-x_{0A}}{Q-r},x_{0B}\right), \hat{N}_{A}^{(W1)}=max\left(r\hat{N}_{B}^{(W1)},x_{0A}\right), 
\end{eqnarray}
where $Q=\left(x_{11B}x_{10A}x_{01A}\right)/\left(x_{11A}x_{10B}x_{01B}\right)$, and $x_{0k}=x_{11k}+x_{10k}+x_{01k}$ is the total numbers of captured individuals from the sub-populations $U_k$ for $k=A,B$.  
In the second model, say Wolter-2, \citet{Wolter90} additionally assumed that the causal independence holds only for the sub-population $U_B$, and the resulting estimates are given by
\begin{eqnarray}\label{WolterM2}
\hat{N}_{B}^{(W2)}=\frac{x_{1\cdot B}x_{\cdot 1B}}{x_{11B}}, \hat{N}_{A}^{(W2)}=max\left(r\hat{N}_{B}^{(W2)},x_{0A}\right).
\end{eqnarray}
See \citet{Wolter90} for more details.

\section{Proposed Model}\label{sec:Bivariate model}
In this section, we first introduce a Bivariate Bernoulli model (BBM), which is useful in measuring the degree of association between capture and recapture attempts. Although the problem can be generalized to a multivariate setup for multiple lists problem, in the present paper we focus our attention to the bivariate version only for DRS. 

In any given population, some individuals are expected to behave independently over the two capture attempts in DRS and dependence in the behavioral responses may exist for rest of the population. Let $\alpha$ be such proportion of individuals for whom behavioral dependence between the List 1 and List 2 exists. To capture this dependency structure, we define a pair ($X_{1h}^{*},X_{2h}^{*}$), which represents the latent capture statuses of the \textit{h}-th individual in the first and second attempts, respectively, for $h=1,2,\ldots,N$. The latent capture status $X_{lh}^{*}$ takes value 1 or 0, denoting the presence or absence of the \textit{h}-th individual in the $l$-th list, for $l=1,2$. Under this setup, for $\alpha$ proportion of individuals, the value of $X_{2h}^{*}$ is same as that of $X_{1h}^{*}$ (i.e. $X_{2h}^{*}=X_{1h}^{*}$). Now, let us define $Y_h$ and $Z_h$, respectively, as the List 1 and List 2 inclusion status of the \textit{h}-th individual belonging to $U$, for $h=1,2,\ldots,N$. Note that $(Y_h,Z_h)$ is manifestation of the latent capture statuses ($X_{1h}^{*},X_{2h}^{*}$) for the \textit{h}-th individual. Therefore, we can formally write the interdependence among the two lists as

$\quad$
\begin{equation}
(Y_h,Z_h) = \begin{cases} (X_{1h}^{*},X_{2h}^{*}) & \mbox{ with prob. } 1-\alpha,\\
(X_{1h}^{*},X_{1h}^{*})  & \mbox{ with prob. } \alpha,
\end{cases}\label{prob-model}
\end{equation}
where $X_{1h}^{*}$s and $X_{2h}^{*}$s are independently and identically distributed Bernoulli random variables with parameters $p_{1}$ and $p_{2}$, respectively. Note that $p_{l}$ refers to the capture probability of a causally independent individual in the \textit{l}-th list. We call this model, given in equation (\ref{prob-model}), as Bivariate Bernoulli model in DRS (BBM-DRS). Now, we denote $Prob(Y=y,Z=z)$ by $p_{yz}$, for $y,z=\{0,1\}$. Thus, based on the parameters involved in the above model, presented in equation (\ref{prob-model}), the cell probabilities associated with DRS (\textit{See} Table \ref{tab:1}) are given by:
\begin{eqnarray}\label{rel}
p_{11} = \alpha p_{1} +(1-\alpha)p_{1}p_{2}, && p_{10} = (1-\alpha)p_{1}(1-p_{2}),\nonumber \\
p_{01} =  (1-\alpha)(1-p_{1})p_{2}, && p_{00} = \alpha(1-p_{1})+(1-\alpha)(1-p_{1})(1-p_{2}).\nonumber
\end{eqnarray}
The corresponding marginal probabilities are given by
\begin{eqnarray}
p_{Y}=p_{1\cdot} = p_{1}, && p_{Z}=p_{\cdot1} = \alpha  p_{1}+(1-\alpha)p_{2},\nonumber
\end{eqnarray}
with $Cov(Y,Z)=\alpha p_{1}(1-p_{1})$.. Note that the proposed Bivariate Bernoulli model incorporates positive dependence between capture status in Lists 1 and 2. In particular, when $\alpha=0$ (i.e. there is no case of causal dependency), our proposed Bivariate Bernoulli model in (\ref{prob-model}) reduces to the $M_t$ model.

\begin{remark}\label{mark_1}
	One can define the proposed BBM-DRS in order to capture negative dependency (or, recapture aversion) by rewriting (\ref{prob-model}) as
	\begin{equation}
	(Y_h,Z_h) = \begin{cases} (X_{1h}^{*},X_{2h}^{*}) & \mbox{ with prob. } 1-\alpha,\\
	(X_{1h}^{*},1-X_{1h}^{*})  & \mbox{ with prob. } \alpha.
	\end{cases}\nonumber\label{prob-model_negative}
	\end{equation}
\end{remark} 



\begin{remark}
	The parameters of BBM-DRS possess easy interpretations with practical significance. The dependence parameter $\alpha$ represents proportion of behaviorally dependent individuals, and $p_{l}$ is the capture probability of an causally independent individual in the \textit{l}-th List, for $i=1,2$.
\end{remark}

\section{Estimation Methodologies}\label{est}
In practice, one can easily consider post-stratification of the entire population into two mutually exclusive and exhaustive sub-populations $U_A$ and $U_B$ as discussed in Section \ref{DRS} (\textit{See} \citet{Wolter90},  \citet{Eisele03} and \citet{Granerod13}). We also assume that for any individual, belonging to $U_A$, the capture status in either of the two lists is independent of the same of an individual belonging to $U_B$. In order to denote the cell counts and the associated probabilities for the $2\times 2$ table obtained under the DRS for the sub-population $U_{k}$, we consider the same notation as mentioned in Table \ref{tab:1}, with an additional suffix $k$ (for example, List 1 capture probability for the sub-population $U_{k}$ is denoted as $p_{1\cdot k}$), for $k=A,B$. Now we consider two different models and propose methodologies for estimation of the associated parameters including the population size $N$$(=N_{A}+N_{B})$, the parameter of primary interest.

\subsection{Model I}\label{Model01}
In this model, we consider the assumption $(S3)$ for the sub-population $U_B$, which implies $p_{11B}=p_{1\cdot B}p_{\cdot1B}$. Therefore, the popular Lincoln-Petersen estimate of $N_{B}$ is given as $\hat{N}_{B}=\textstyle \left(\frac{x_{1\cdot B}x_{\cdot1B}}{x_{11B}}\right)$. In order to incorporate the behavioural dependency present in the sub-population $U_{A}$, we consider BBM-DRS as described in Subsection \ref{sec:Bivariate model}, which consists of four parameters with $p_{1}=p_{1 A}, p_{2}=p_{2 A}, \alpha=\alpha_{A}$, and $N=N_A$. In addition to $(S3)$, we consider the following assumption:\\

$(S4)$ Initial (List 1) capture probabilities for the individuals belonging to both the sub-populations $U_A$ and $U_B$ are the same (i.e. $p_{1\cdot A}=p_{1\cdot B}=p_1,say$).\\

The assumption $(S4)$ ensures estimability of the model parameters. Note that List 1 is prepared before List 2 and hence, List 2 capture probabilities for different sub-populations may differ due behavioral dependence, if exists. Also, it is quite reasonable to consider the same List 1 capture probability for different sub-populations when possibly there is no prejudice. Similar assumption has been considered by several authors in the past \citep{Bell93}. Under similar setup, \citet{Wolter90} proposed estimate of $N_{B}$ based on $M_{t}$ model and the estimate of $N_{A}$ using the available knowledge on the ratio of the sub-population sizes (e.g. sex-ratio). As discussed before, the availability of reliable estimate of this ratio remains a practical challenge (\textit{See Section \ref{realdata}}). As mentioned before, $N_B$ is estimated assuming causal independence, and hence, one needs to find the estimate of $N_A$ in order to estimate the population size $N$. Since $\alpha_{A}$ can be interpreted as the proportion of behaviorally dependent individuals, its estimation may provide interesting insight of the capture-recapture mechanism.

First we consider method of moments estimation of the parameters associated with the proposed Model I. Note that the method of moments estimate (MME) of $N_{B}$ is same as the Lincoln-Petersen estimate $\hat{N}_{B}=\textstyle \left(\frac{x_{1\cdot B}x_{\cdot1B}}{x_{11B}}\right)$, and the MMEs of $p_{1B}$ and $p_{2B}$ are given as $\hat{p}_{1B}=\frac{x_{11B}}{x_{\cdot1B}}$ and $\hat{p}_{2B}=\frac{x_{11B}}{x_{1\cdot B}}$, respectively. Using the assumption $(S4)$, the estimate of $p_{1 A}$ is given by $\hat{p}_{1A}=\hat{p}_{1}=\frac{x_{11B}}{x_{\cdot1B}}$. Now, equating the expected and observed number of cell counts in the $2\times 2$ table obtained under the DRS (Table \ref{tab:1}) for the sub-population $U_{A}$, we get
\begin{eqnarray}\label{MME_MI}
N_{A}p_{11A} = x_{11A}\label{relation_1}, & N_{A}p_{10A} = x_{10A}\mbox{ and}\label{relation_2} & N_{A}p_{01A} = x_{01A}\label{relation_3},
\end{eqnarray}
which involve three unknown parameters $N_A, p_{2 A}$, and $\alpha_{A}$. Solving these equations in (\ref{MME_MI}), the MMEs of the model parameters are obtained as\\
\begin{eqnarray}
\hat{N}_{A}^{(1)} &=&\displaystyle \frac{x_{1\cdot A}x_{\cdot1B}}{x_{11B}},\label{eqn_est_N}\nonumber\\
\hat{p}_{2A}^{(1)} & = & \frac{x_{01A}x_{11B}}{x_{10A}x_{01B}+x_{01A}x_{11B}},\label{eqn_est_p2}\nonumber\\
\hat{\alpha}_{A}^{(1)} & = &\min\left\{ \max\left\{0,\frac{x_{\cdot1A}}{x_{1\cdot A}}-\frac{x_{01A}x_{\cdot1B}}{x_{01B}x_{1\cdot A}}\right\},1\right\}.\label{eqn_est_alpha}\nonumber
\end{eqnarray}
The detailed derivation for finding the above mentioned MMEs are provided in the \textit{Appendix}.

A classical approach for estimating $N$ from an incomplete $2\times2$ cross-classified data structure, is based on \textit{likelihood theory}, where the data (i.e. all observed cell counts in Table \ref{tab:1}) follow a multinomial distribution with index parameter $N$ and the associated cell probabilities $\{p_{ijk}:i,j=0,1;i=j\neq0, k=A, B\}$ \citep{Sanathanan72}. 
Therefore, using the relations between the cell probabilities $\{p_{ijk}\}$ and
$\theta_1=(N_A,N_B,\alpha_{A},p_1,p_{2A},p_{2B})$, as provided in Section \ref{sec:Bivariate model}, the likelihood function of $\theta_{1}$ is given by
\begin{eqnarray}\label{L_Model_I}
L(\theta_1|\underline{\textbf{x}}_A,\underline{\textbf{x}}_B) & \propto & \frac{N_A! N_B!}{(N_A-x_{0A})!(N_B-x_{0B})!}   [\alpha_{A} p_1+(1-\alpha_{A})p_1p_{2A}]^{x_{11A}}\nonumber\\
&&\times
p_1^{({x_{10A}+x_{11B}+x_{10B}})} 
(1-p_{1})^{(x_{01A}+N_{B}-x_{11B}-x_{10B})}p_{2A}^{x_{01A}}\nonumber\\
&&\times
p_{2B}^{(x_{11B}+x_{01B})}(1-p_{2A})^{x_{10A}}(1-p_{2B})^{(N_{B}-x_{11B}-x_{01B})}
(1-\alpha_{A})^{(x_{10A}+x_{01A})}\nonumber\\
&&\times[\alpha_{A}(1-p_1)+(1-\alpha_{A})(1-p_1)(1-p_{2A})]^{(N_A-x_{0A})},
\end{eqnarray}
where $\underline{\textbf{x}}_k=\left(x_{11k},x_{10k},x_{01k}\right)$, $x_{0k}=x_{11k}+x_{10k}+x_{01k}$, for $k=A,B$. However, explicit solution for maximum likelihood estimate (MLE) of $\theta_1$ is not possible. The Newton-Raphson method can be used to maximize the log-likelihood in order to estimate $\theta_1$, assuming $N_{A}$ and $N_{B}$ as continuous parameters. Alternatively, any standard software package equipped with general purpose optimization (e.g., \textit{optim} in the package R) can be used. Note that the log-likelihood function involves $\ln(N_{A}!)$, which may create computational difficulty for large values of $N_{A}$. In order to avoid such issues we approximate $\ln(N_{A}!)$ as $N_{A} \ln(N_{A})-N_{A}+\frac{1}{2}\ln(2\pi N_{A})$ \citep[p. 45]{Wells86}.

\begin{remark}
	The above likelihood function (\ref{L_Model_I}) can be simplified using Stirling's approximation of $\ln(N_{A}!) \approx N_{A} \ln(N_{A})-N_{A}$ \citep[p. 138-140]{Whittaker67}, and obtain closed form expression of the MLEs. Interestingly, the MLEs for all the parameters are exactly equal to the respective MMEs.
\end{remark}

\begin{remark}\label{Remark_sex}
	If the ratio of the sub-population sizes (e.g. sex-ratio for male-female stratification) $r$ is known, one can easily incorporate such information in the likelihood function (\ref{L_Model_I}) taking $N_{B}=r^{-1} N_{A}$.
\end{remark}

\subsection{Model II}\label{Model02}
In Model II, we relax the assumption $(S3)$ and the BBM-DRS is considered for both the sub-populations $U_A$ and $U_B$ with parameters $p_{1}=p_{1 k}$, $p_{2}=p_{2 k}$, $\alpha=\alpha_k$, and $N=N_k$, for $k=A,B$. Similar to Model I, we consider the assumption $(S4)$ (i.e. $p_{1 A}=p_{1 B}=p_{1}$, say) and additionally we assume $\alpha_A=\alpha_B=\alpha_{0}$, say, which ensures estimability of Model II. Under similar setup, \citet{Wolter90} proposed estimates of $N_{A}$ and $N_{B}$ using the ratio of the sub-population sizes. As discussed before, reliable estimate of this ratio is not available in most of the cases.

We first consider the method of moments for estimating the parameters associated with the Model II. We equate the expected and observed cell counts from the $2\times 2$ tables obtained under the DRS involving six parameters $N_A,N_B, p_{1},p_{2A},p_{2B},\alpha_{0}$ and find the following MMEs as

\begin{eqnarray}
\hat{p}_{2A}^{(2)}&=&\frac{x_{01B}(x_{1\cdot A}x_{10B}-x_{1\cdot B}x_{10A})}{x_{1\cdot B}(x_{01A}x_{10B}-x_{10A}x_{01B})},\nonumber\\
\hat{p}_{2B}^{(2)}&=&\frac{x_{01A}(x_{1\cdot A}x_{10B}-x_{1\cdot B}x_{10A})}{x_{1\cdot A}(x_{01A}x_{10B}-x_{10A}x_{01B})},\nonumber\\
\hat{\alpha}_{0}^{(2)}&=&1-\frac{x_{10A}}{x_{1\cdot A}}\frac{1}{1-\hat{p}_{2A}},\nonumber\\
\hat{p}_1^{(2)}&=&\frac{1}{1+\frac{x_{01A}}{x_{10A}}\left(\frac{1}{\hat{p}_{2A}}-1\right)},\nonumber\\
\hat{N}_A^{(2)}&=&\frac{x_{1\cdot A}}{\hat{p}_1},\nonumber\\
\hat{N}_B^{(2)}&=&\frac{x_{1\cdot B}}{\hat{p}_1}.\nonumber
\end{eqnarray}
The derivation for finding the above mentioned MMEs is similar to that of Model I. See \textit{Appendix} for more details. In some cases $\hat{p}_{2A}>\frac{x_{01A}}{x_{01A}-x_{10A}}$, and hence, the estimates for $p_1 $, $N_A$ and $N_B$ become negative, as in \citet{Wolter90}. Such issues with MME has been discussed in the literature (\textit{See} \citet[p. 2092-2098]{Bowman98} for more details). Therefore, it is not advisable to use MME for the proposed Model II and one should prefer the maximum likelihood estimates as provided below. 

Using the relations between the cell probabilities $\{p_{ijk}\}$ and
$\theta_2=(N_A,N_B,\alpha_{0},p_1,p_{2A},p_{2B})$, as provided in Section \ref{sec:Bivariate model}, the likelihood function of $\theta_{2}$ is given by
\begin{eqnarray}\label{L_Model_II}
L(\theta_2|\underline{\textbf{x}}_A,\underline{\textbf{x}}_B) & \propto & \frac{N_A! N_B!}{(N_A-x_{0A})!(N_B-x_{0B})!}   [\alpha_{0} p_1+(1-\alpha_{0})p_1p_{2A}]^{x_{11A}}\nonumber\\
&&\times
[\alpha_{0} p_1+(1-\alpha_{0})p_1p_{2B}]^{x_{11B}} p_{1}^{(x_{10A}+x_{10B})}(1-p_{1})^{(x_{01A}+x_{01B})}\nonumber\\
&&\times
p_{2A}^{x_{01A}}p_{2B}^{x_{01B}}(1-p_{2A})^{x_{10A}}(1-p_{2B})^{x_{10B}}
(1-\alpha_{0})^{(x_{10A}+x_{01A}+x_{10B}+x_{01B})}\nonumber\\
&&\times[\alpha_{0}(1-p_1)+(1-\alpha_{0})(1-p_1)(1-p_{2A})]^{(N_A-x_{0A})}\nonumber\\
&&\times[\alpha_{0}(1-p_1)+(1-\alpha_{0})(1-p_1)(1-p_{2B})]^{(N_B-x_{0B})},
\end{eqnarray}
where $\underline{\textbf{x}}_k=\left(x_{11k},x_{10k},x_{01k}\right)$, $x_{0k}=x_{11k}+x_{10k}+x_{01k}$, for $k=A,B$. Since, the explicit solution for MLE of $\theta_2$ cannot be obtained, same computational strategy is followed here as in the case of Model I. As remarked in Subsection \ref{Model01}, here also one can consider the same reparameterization $N_{B}=r^{-1} N_{A}$ in the likelihood function (\ref{L_Model_II}), if the ratio of the sub-population sizes $r$ is known.

\section{Simulation Study}\label{simu}
In this section, the performance of the proposed estimators are thoroughly investigated based on simulation study and compared with the existing competitors. For this purpose, we consider six trial populations, denoted by $P1,\ldots, P6$, with the choices of capture probabilities $\left(p_{1\cdot k}, p_{\cdot1k}\right)=\left\{(0.60, 0.80), (0.60, 0.70), (0.80, 0.55), (0.80, 0.70), (0.50, 0.75), (0.50, 0.60)\right\}$, respectively, for $k=\{A,B\}$, with $(N_{A},N_{B})=(240,200)$ and $(1200,1000)$. We present the simulation study in two fold. Firstly, we consider the ratio of the sub-population sizes $r$ ($=N_A/N_B$) is unknown and compare the performance of our proposed estimators with the Nour's \citep{Nour82} estimator given by (\ref{Nour}). As discussed before, the estimators, (\ref{WolterM1}) and (\ref{WolterM2}), proposed by \citet{Wolter90} are not applicable here. Secondly, we consider $r$ is known and the Wolter's \cite{Wolter90} estimators are compared with the proposed estimators. It is important to note that the Nour's \citep{Nour82} method is unable to incorporate the knowledge on $r$.   

First, we generate 1000 data sets from Model I for each of the six said trial populations $P1-P6$ with $\alpha_A=0.4,0.8$. As the LP estimator of $N_B$ produces efficient results under the causal independence assumption (S3) for large or moderately large samples, our primary interest in Model I lies in the estimate of $N_{A}$ based on MME and MLE. Final estimate of $N_{A}$ is obtained by averaging the estimates over $1000$ replications. To compare the performance of the estimators, we compute relative bias (RB) and relative root mean square error (RRMSE) using the following formula:
\begin{eqnarray}
RB & = & \frac{1}{N_{A}}\left[\left(\frac{1}{1000}\sum_{j=1}^{1000}\hat{N}_A^{(j)}\right)-N_A\right],\nonumber\\
RRMSE & = & \frac{1}{N_{A}}\left[\frac{1}{1000}\sum_{j=1}^{1000}\left(\hat{N}_A^{(j)}-N_A\right)^2\right]^{1/2}. \nonumber
\end{eqnarray}
In the capture-recapture setting, point estimators of population size are commonly possess positively skewed distributions \citep{Yang10}. Therefore, we obtain $95\%$ confidence interval (C.I.) for $N_A$ based on the log-transformation method, discussed in \citet{Chao87} and \citet{Yang05}. In this method, $log(\hat{N}_A-x_{0A})$ is approximately treated as normal variate and that gives $95\%$ confidence interval as 
\[\left[x_{0A}+\left(\hat{N}_A-x_{0A}\right)\left/\right.C,\ x_{0A}+\left(\hat{N}_A-x_{0A}\right)C\right],\]
where $C=exp\left\{1.96\left[log\left(1+\hat{\sigma}_{\hat{N}_A}^2/(\hat{N}_A-x_{0A})^2\right)\right]^{1/2}\right\}$, and $\hat{\sigma}_{\hat{N}_A}^2$ is the estimate of the variance of $\hat{N}_A$. For each of the 1000 replications, $\hat{\sigma}_{\hat{N}_A}$ is computed using parametric bootstrap method based on 1000 bootstrap samples. Length of the $95\%$ confidence interval (LCI) as well as its coverage probability (CP) are computed following the methods discussed in \citet{Yang10} and \citet{Chatterjee18}. First, we need to compare the CPs of each of the estimators to see which one performs the best. Further, we need to compare the LCIs when coverage probabilities (CPs) are found to be similar \citep{Yang10}. Note that Nour's \citep{Nour82} estimator is not model-based and corresponding CP and LCI cannot be obtained by the aforementioned parametric bootstrap method. The results are presented in Tables \ref{tab:2a} and \ref{tab:2b} for true population size $N_{A}=240$ and $1200$, respectively. 	

From Table \ref{tab:2a}, it is observed that both the proposed estimators (MME and MLE) of $N_{A}$ outperform the Nour's \citep{Nour82} estimator in terms of RB and RRMSE. One can also observe that the RB and RRMSE of the MLE are smaller compared to that of the MME. Interestingly, the performance of the MLE and MME are comparable with respect to CP and LCI. As expected, both RB and RRMSE of the proposed estimators decrease as the population size $N_A$ increases.

Next, we generate data from Model II considering the same trial populations $P1-P6$ along with common dependence parameter $\alpha_{0}=0.4,0.8$. Similar to the case of Model I, we obtain RB, RRMSSE, CP, and LCI for estimators of both $N_A$ and $N_B$, and the results are presented in Tables \ref{tab:3a} and \ref{tab:3b}. As discussed before, the proposed MME from the Model II is often found to be negative; hence, these estimator has not been considered for this simulation study. It is clear from the results presented in Table \ref{tab:3a} that the performance of the proposed MLE under Model II is significantly better than that of \citet{Nour82} both in terms of RB and RRMSE. Nour's \citep{Nour82} estimator underestimates the $N_A$ and $N_B$, where as the biases incurred by our proposed MLE are negligible for both the sub-population sizes. The results from Table \ref{tab:3a} indicate that the interval estimates based on the proposed MLE performs efficiently both in terms of CP as well as LCI. As expected, the RB and RRMSE of the MLE decreases as the population sizes $N_A$ and $N_B$ increases. 

As mentioned in Remark \ref{Remark_sex}, information on the ratio of the sub-population size $r$, if available, can be incorporated in our proposed likelihood based estimate. It is important to note that the estimate of the ratio of sub-population sizes $r$ may be available for large population based on previous studies \citep{Wolter90}. For example, in a census coverage study, estimate of the sex-ratio may be available from a past demographic analysis of the population under consideration\citep{Robinson93}. Therefore, assuming $r$ to be known, we presented this analysis only for the large populations, that is for $(N_A,N_B)=(1200,1000)$.
The performance of our proposed estimator under Model I (Model II) is compared with the estimator of Wolter-2 (Wolter-1) and the results are presented in Table \ref{tab:4a} (Table \ref{tab:4b}). It is clearly seen that the proposed estimator is superior than Wolter's estimators with respect to RB and RRMSE. Moreover, our models produce far better CPs of its 95\% CIs than that of the models proposed by \citet{Wolter90} for both the choices of $\alpha_A$ or $\alpha_0$. The resulting CIs from Wolter-2 has shorter lengths than that of our Model I, however, Wolter-1 exhibits much more wider confidence intervals compared to the proposed Model II.  Similar results are also observed (not reported here) for $(N_A,N_B)=(240,200)$. 

\begin{table}[ht]
	\small
	\centering
	\caption{Summary results on the estimators of $N_A$ under the simulation model  Model I with $N_A=240$ and the ratio of the sub-population sizes ($r$) is unknown.}
	\begin{tabular}{|cclrccc|}
		\hline
		\multicolumn{7}{|c|}{}\\
		Population & $\alpha_A$ & \multicolumn{1}{c}{Method} & \multicolumn{1}{c}{RB} & RRMSE & CP($\%$)& LCI\\
		\hline
		\hline
		\multicolumn{7}{|c|}{}\\ 
		P1 & 0.4 & MLE & 0.0370 & 0.0632 & 90 & 84.47 \\
		&     & MME & 0.0684 & 0.0843 & 99 & 107.75 \\
		&     & Nour & 0.1669 & 0.1703  & - & - \\  
		& 0.8 & MLE & 0.0361 & 0.0628 & 99.5 & 98.84 \\
		&     & MME & 0.0650 & 0.0813 & 99 & 99.55 \\
		&     & Nour & 0.3377 & 0.3389 & - & - \\
		\multicolumn{7}{|c|}{}\\ 
		P2 & 0.4 & MLE & 0.0420 & 0.0647 & 88.5 & 95.53 \\
		&     & MME & 0.0747 & 0.0940 & 96.5 & 111.58 \\
		&     & Nour & 0.1628 & 0.1660 & - & - \\  
		& 0.8 & MLE & 0.0413 & 0.0660 & 100 & 99.34 \\
		&     & MME & 0.0719 & 0.0898 & 98.5 & 102.39 \\
		&     & Nour & 0.3346 & 0.3358 & - & - \\
		\multicolumn{7}{|c|}{}\\ 
		P3 & 0.4 & MLE & 0.0401 & 0.0608 & 85 & 45.62 \\
		&     & MME & 0.0497 & 0.0634 & 94.5 & 76.87 \\
		&     & Nour & 0.0847 & 0.0886 & - & - \\  
		& 0.8 & MLE & 0.0367 & 0.0911 & 92.7 & 80.89 \\
		&     & MME & 0.0475 & 0.0594 & 98 & 71.56 \\
		&     & Nour & 0.1701 & 0.1717 & - & - \\
		\multicolumn{7}{|c|}{}\\ 
		P4 & 0.4 & MLE & 0.0318 & 0.0495 & 90 & 43.18 \\
		&     & MME & 0.0449 & 0.0582 & 25 & 73.64 \\
		&     & Nour & 0.0822 & 0.0861 & -  &  - \\  
		& 0.8 & MLE & 0.0291 & 0.0644 & 88.89 & 77.53 \\
		&     & MME & 0.0400 & 0.0519 & 97.5 & 68.79 \\
		&     & Nour & 0.1687 & 0.1703 & - & - \\
		\multicolumn{7}{|c|}{}\\  
		P5 & 0.4 & MLE & 0.0391 & 0.0701 & 92 & 102.32 \\
		&     & MME & 0.0872 & 0.1093 & 79 & 131.05 \\
		&     & Nour & 0.2065 & 0.2100 & - &  -\\  
		& 0.8 & MLE & 0.0401 & 0.0724 & 99 & 136.63 \\
		&     & MME & 0.0791 & 0.1040 & 97.5 & 127.81 \\
		&     & Nour & 0.4174 & 0.4187 & - & - \\
		\multicolumn{7}{|c|}{}\\  
		P6 & 0.4 & MLE & 0.0377 & 0.0754 & 90 & 101.78 \\
		&     & MME & 0.0877 & 0.1131 & 96.5 & 139.50 \\
		&     & Nour & 0.2135 & 0.2169 & - & - \\  
		& 0.8 & MLE & 0.0487 & 0.0681 & 99 & 141.74 \\
		&     & MME & 0.0868 & 0.1099 & 98 & 136.50 \\
		&     & Nour & 0.4223 & 0.4236 & - & - \\		      
		\hline
	\end{tabular}
	\label{tab:2a}
\end{table}

\begin{table}[ht]
	\small
	\centering
	\caption{Summary results on the estimators of $N_A$ under the simulation model Model I with $N_A=1200$ and the ratio of the sub-population sizes ($r$) is unknown.}
	\begin{tabular}{|cclrccc|}
		\hline
		\hline
		\multicolumn{7}{|c|}{}\\
		Population & $\alpha_A$ & \multicolumn{1}{c}{Method} & \multicolumn{1}{c}{RB} & RRMSE & CP($\%$)& LCI\\
		\hline
		\hline
		\multicolumn{7}{|c|}{}\\ 
		P1 & 0.4 & MLE & 0.0002 & 0.0153 & 99 & 295.07 \\
		&     & MME & -0.0002 & 0.0381 & 99 & 219.52 \\
		&     & Nour & -0.1661 & 0.1669 & - & - \\  
		& 0.8 & MLE & 0.0005 & 0.0182 & 99 & 160.94 \\
		&     & MME & -0.0019 & 0.0382 & 98 & 221.31 \\
		&     & Nour & -0.3371 & 0.3373 & - & - \\
		\multicolumn{7}{|c|}{}\\ 
		P2 & 0.4 & MLE & 0.0017 & 0.0138 & 99.5 & 412.65 \\
		&     & MME & 0.0035 & 0.0388 & 98.5 & 223.56 \\
		&     & Nour & -0.1626 & 0.1633 & - & - \\  
		& 0.8 & MLE & 0.0002 & 0.0161 & 99.5 & 165.51 \\
		&     & MME & 0.0019 & 0.0388 & 99.5 & 227.80 \\
		&     & Nour & -0.3339 & 0.3342 & - & - \\
		\multicolumn{7}{|c|}{}\\         
		P3 & 0.4 & MLE & 0.0027 & 0.0144 & 99.5 & 158.39 \\
		&     & MME & 0.0021 & 0.0262 & 98 & 154.01  \\
		&     & Nour & -0.0851 & 0.0860 & - & - \\  
		& 0.8 & MLE & 0.0017 & 0.0125 & 100 & 182.97\\
		&     & MME & 0.0008 & 0.0264 & 98.5 & 159.08 \\
		&     & Nour & -0.1723 & 0.1726 & - & - \\
		\multicolumn{7}{|c|}{}\\ 
		P4 & 0.4 & MLE & 0.0009 & 0.0089 & 100 & 233.09 \\
		&     & MME & 0.0025 & 0.0265 & 97 & 151.14\\
		&     & Nour & -0.0825 & 0.0834  & - & - \\  
		& 0.8 & MLE & 0.0012 & 0.0089 & 100 & 291.15  \\
		&     & MME & 0.0011 & 0.0261 & 97.5 & 152.76 \\
		&     & Nour & -0.1705 & 0.1709 & - & - \\
		\multicolumn{7}{|c|}{}\\
		P5 & 0.4 & MLE & 0.0008 & 0.0098 & 100 & 226.03 \\
		&     & MME &  -0.0009 & 0.0466 & 98.5 & 263.76 \\
		&     & Nour &  -0.2064 & 0.2071  & - & - \\  
		& 0.8 & MLE & 0.0002 & 0.0180 & 100 & 286.80 \\
		&     & MME & -0.0006 & 0.0448 & 99 & 265.13 \\
		&     & Nour & -0.4208 & 0.4210 & - & - \\
		\multicolumn{7}{|c|}{}\\ 
		P6 & 0.4 & MLE & -0.0002 & 0.0195 & 99.5 & 309.30 \\
		&     & MME & 0.0026 & 0.0505 & 99 & 276.84 \\
		&     & Nour & -0.2128 & 0.2135 & - & - \\  
		& 0.8 & MLE & 0.0002 & 0.0218 & 99.5 & 378.88  \\
		&     & MME & 0.0030 & 0.0508 & 99.5 & 277.64 \\
		&     & Nour & -0.4248 & 0.4250 & - & - \\
		\hline
	\end{tabular}
	\label{tab:2b}
\end{table}

\begin{table}[ht]
	\scriptsize
	\centering
	\caption{Summary results on the estimators of $N_A$ \& $N_B$ under the simulation model Model II with $(N_A, N_B)=(240,200)$ and the ratio of the sub-population sizes ($r$) is unknown.}
	\begin{tabular}{|cclrccc|}
		\hline
		\multicolumn{7}{|c|}{}\\
		Population & $\alpha_0$ & \multicolumn{1}{c}{Method} & \multicolumn{1}{c}{RB} & RRMSE & CP($\%$)& LCI\\
		\hline
		\hline
		\multicolumn{7}{|c|}{}\\
		\multicolumn{7}{|c|}{Results on estimators of $N_A$}\\ 
		\hline
		P1 & 0.4 & MLE  & 0.0028 & 0.0265 & 96.5 & 27.84 \\
		&     & Nour & -0.1692  & 0.1728  & - & - \\  
		& 0.8 & MLE & 0.0073 & 0.0401 & 96 & 44.36  \\
		&     & Nour  & -0.3382 & 0.3398  & - & - \\  
		\multicolumn{7}{|c|}{}\\ 
		P2 & 0.4 & MLE  & 0.0034  & 0.0300 & 95.5 & 27.22  \\
		&     & Nour & -0.1659 &  0.1691 & - & - \\  
		& 0.8 & MLE & 0.0067  & 0.0424 & 96 & 42.23  \\
		&     & Nour  & -0.3350 & 0.3365  & - & - \\
		\multicolumn{7}{|c|}{}\\ 
		P3 & 0.4 & MLE  & 0.0108  & 0.0425 & 92.5 & 40.81  \\
		&     & Nour & -0.0883 & 0.0920  & - & - \\  
		& 0.8 & MLE & 0.0288  & 0.0742 & 86.5 & 57.93  \\
		&     & Nour  & -0.1721 & 0.1741  & - & - \\
		\multicolumn{7}{|c|}{}\\ 
		P4 & 0.4 & MLE  & 0.0072 & 0.0379 & 95 & 37.11  \\
		&     & Nour & -0.0857 & 0.0893  & - & - \\  
		& 0.8 & MLE & 0.0208  & 0.0603 & 92 & 55.67  \\
		&     & Nour  & -0.1704 & 0.1724  & - & - \\
		\multicolumn{7}{|c|}{}\\ 
		P5 & 0.4 & MLE  & 0.0040 & 0.0291 & 95.5 & 31.02  \\
		&     & Nour & -0.2066 &  0.2103 & - & - \\  
		& 0.8 & MLE & 0.0126 & 0.0473 & 96.5 & 51.26 \\
		&     & Nour  &  -0.4188 & 0.4199  & - & - \\
		\multicolumn{7}{|c|}{}\\ 
		P6 & 0.4 & MLE  & 0.0053 & 0.0328 & 93.5 & 32.54  \\
		&     & Nour & -0.2117 & 0.2157  & - & - \\  
		& 0.8 & MLE & 0.0087  & 0.0419 & 97 & 48.92  \\
		&     & Nour  & -0.4225  & 0.4236 & - & - \\ 
		\hline 
		\multicolumn{7}{|c|}{}\\
		\multicolumn{7}{|c|}{Results on estimators of $N_B$}\\ 
		\hline
		P1 & 0.4 & MLE  & 0.0302 & 0.0374 & 94.5 & 30.61  \\
		&     & Nour & -0.1603 & 0.1639  & - & - \\  
		& 0.8 & MLE & 0.0103 & 0.0505 & 95 & 44.01  \\
		&     & Nour  & -0.3305 & 0.3321  & - & - \\  
		\multicolumn{7}{|c|}{}\\ 
		P2 & 0.4 & MLE  & 0.0076 & 0.0386  & 93 & 29.65  \\
		&     & Nour & -0.1656 & 0.1699   & - & - \\  
		& 0.8 & MLE & 0.0095 & 0.0491 & 97 & 44.00 \\
		&     & Nour  & -0.3367 & 0.3383  & - & - \\
		\multicolumn{7}{|c|}{}\\ 
		P3 & 0.4 & MLE  & 0.0166 & 0.0501 & 93  & 38.01 \\
		&     & Nour &  -0.0854 & 0.0908   & - & - \\  
		& 0.8 & MLE & 0.0314 & 0.0761  & 88.5 & 52.43 \\
		&     & Nour  & -0.1730 & 0.1746   & - & - \\
		\multicolumn{7}{|c|}{}\\ 
		P4 & 0.4 & MLE  & 0.0120 & 0.0407  & 91.5 & 32.28 \\
		&     & Nour &  -0.0795 & 0.0839   & - & - \\  
		& 0.8 & MLE &  0.0229 & 0.0647 & 92.5 & 49.06 \\
		&     & Nour  & -0.1678 & 0.1695   & - & - \\
		\multicolumn{7}{|c|}{}\\ 
		P5 & 0.4 & MLE  &  0.0042 & 0.0438  & 95.5 & 35.63 \\
		&     & Nour &  -0.2014 & 0.2059   & - & - \\  
		& 0.8 & MLE & 0.0120 & 0.0656  & 93 & 52.44  \\
		&     & Nour  & -0.41283 & 0.4144  & - & - \\
		\multicolumn{7}{|c|}{}\\ 
		P6 & 0.4 & MLE  & 0.0029 & 0.0456 & 97 & 37.18 \\
		&     & Nour & -0.2223 & 0.2262  & - & - \\  
		& 0.8 & MLE &  0.0101 & 0.0604  & 95.5 & 51.97 \\
		&     & Nour  &  -0.42544 & 0.4270   & - & - \\        
		\hline
	\end{tabular}
	\label{tab:3a}
\end{table}

\begin{table}[ht]
	\scriptsize
	\centering
	\caption{Summary results on the estimators of $N_A$ \& $N_B$ under the simulated model Model II with $(N_A, N_B)=(1200,1000)$ and the ratio of the sub-population sizes ($r$) is unknown.}
	\begin{tabular}{|cclrccc|}
		\hline
		\multicolumn{7}{|c|}{}\\
		Population & $\alpha_0$ & \multicolumn{1}{c}{Method} & \multicolumn{1}{c}{RB} & RRMSE & CP($\%$)& LCI\\
		\hline
		\hline
		\multicolumn{7}{|c|}{}\\
		\multicolumn{7}{|c|}{Results on estimators of $N_A$}\\ 
		\hline
		P1 & 0.4 & MLE  & 0.0004 & 0.0097 & 96 & 46.14 \\
		&     & Nour & -0.1648 & 0.1654  & - & - \\  
		& 0.8 & MLE & 0.0001 & 0.0127 & 93.5 & 61.19  \\
		&     & Nour  & -0.3371 & 0.3374  & - & - \\  
		\multicolumn{7}{|c|}{}\\ 
		P2 & 0.4 & MLE  & 0.0008 & 0.0111 & 95 & 52.49  \\
		&     & Nour & -0.1615 & 0.1621 & - & - \\  
		& 0.8 & MLE & 0.0001 & 0.0135 & 93.5 & 64.33 \\
		&     & Nour  & -0.3337 & 0.3341  & - & - \\
		\multicolumn{7}{|c|}{}\\ 
		P3 & 0.4 & MLE  & 0.0001 & 0.0069 & 93.5 & 33.13  \\
		&     & Nour & -0.0849 & 0.0856  & - & - \\  
		& 0.8 & MLE & 0.0010 & 0.0088 & 94.5 & 41.02  \\
		&     & Nour  & -0.1709 & 0.1714 & - & - \\
		\multicolumn{7}{|c|}{}\\ 
		P4 & 0.4 & MLE  & 0.0001 & 0.0063 & 94.5 & 30.10  \\
		&     & Nour & -0.0825 & 0.0831  & - & - \\  
		& 0.8 & MLE & 0.0009 & 0.0084 & 95 & 38.98 \\
		&     & Nour  & -0.1691 & 0.1695  & - & - \\
		\multicolumn{7}{|c|}{}\\ 
		P5 & 0.4 & MLE  & -0.0011 & 0.0121 & 95 & 57.61  \\
		&     & Nour & -0.2084 & 0.2091 & - & - \\  
		& 0.8 & MLE & 0.0001 & 0.0157 & 94.5 & 75.60 \\
		&     & Nour  & -0.4210 & 0.4212  & - & - \\
		\multicolumn{7}{|c|}{}\\ 
		P6 & 0.4 & MLE  & -0.0012 & 0.0139 & 94.5 & 66.93 \\
		&     & Nour & -0.2149 & 0.2157  & - & - \\  
		& 0.8 & MLE & 0.00021 & 0.0168 & 95 & 80.66  \\
		&     & Nour  & -0.4253 & 0.4255 & - & - \\  
		\hline
		\multicolumn{7}{|c|}{}\\
		\multicolumn{7}{|c|}{Results on estimators of $N_B$}\\ 
		\hline
		P1 & 0.4 & MLE  & -0.0001 & 0.0150 & 95 & 58.85  \\
		&     & Nour & -0.1603 & 0.1611  & - & - \\  
		& 0.8 & MLE & 0.0003 & 0.0187 & 95.5 & 74.76  \\
		&     & Nour  & -0.3304 & 0.3307  & - & - \\  
		\multicolumn{7}{|c|}{}\\ 
		P2 & 0.4 & MLE  & -0.0005 & 0.0152  & 95 & 60.38  \\
		&     & Nour & -0.1664 & 0.1673   & - & - \\  
		& 0.8 & MLE & 0.0003 & 0.0192 & 96.5 & 75.81 \\
		&     & Nour  & -0.3367 & 0.3370  & - & - \\
		\multicolumn{7}{|c|}{}\\ 
		P3 & 0.4 & MLE  & 0.0005  & 0.0097 & 94 & 38.60 \\
		&     & Nour & -0.0879 & 0.0891   & - & - \\  
		& 0.8 & MLE & -0.0008 & 0.0123 & 95 & 48.36 \\
		&     & Nour  & -0.1741 & 0.1745   & - & - \\
		\multicolumn{7}{|c|}{}\\ 
		P4 & 0.4 & MLE  & 0.0005 & 0.0092  & 94.5 & 36.81 \\
		&     & Nour &  -0.0809 & 0.0819  & - & - \\  
		& 0.8 & MLE &  0.0009 & 0.0121 & 94.5 & 46.92 \\
		&     & Nour  & -0.1689 & 0.1693 & - & - \\
		\multicolumn{7}{|c|}{}\\ 
		P5 & 0.4 & MLE  &  0.0023 & 0.0184  & 95 & 71.12 \\
		&     & Nour &  -0.2012 & 0.2020   & - & - \\  
		& 0.8 & MLE & 0.0004 & 0.0231 & 95 & 91.65 \\
		&     & Nour  & -0.4165 & 0.4168  & - & - \\
		\multicolumn{7}{|c|}{}\\ 
		P6 & 0.4 & MLE  & 0.0023 & 0.0196 & 96.5 & 77.21 \\
		&     & Nour & -0.2198 & 0.2207 & - & - \\  
		& 0.8 & MLE &  0.0003 & 0.0240  & 96 & 95.88 \\
		&     & Nour  & -0.4289 & 0.4293   & - & - \\        
		\hline
	\end{tabular}
	\label{tab:3b}
\end{table}

\begin{table}[ht]
	\small
	\centering
	\caption{Summary results on the estimators of $N_A$ under the simulation model Model I with $N_A=1200$ and the ratio of the sub-population sizes ($r=1.2$) is known.}
	\begin{tabular}{|cclrccc|}
		\hline
		\multicolumn{7}{|c|}{}\\
		Population & $\alpha_A$ & \multicolumn{1}{c}{Method} & \multicolumn{1}{c}{RB} & RRMSE & CP($\%$)& LCI\\
		\hline
		\hline
		\multicolumn{7}{|c|}{}\\ 
		P1 & 0.4 & MLE & -0.0010 & 0.0111 & 100 & 149.35 \\
		&     & Wolter-2 & -0.0013 & 0.0131 & 54 & 23.31 \\
		& 0.8 & MLE & -0.0008 & 0.0121 & 100 & 206.59 \\
		&     & Wolter-2 & -0.0013 & 0.0131 & 21 & 9.16 \\
		\multicolumn{7}{|c|}{}\\ 
		P2 & 0.4 & MLE & 0.0014 & 0.0151 & 100 & 152.80 \\
		&     & Wolter-2 & 0.0007 & 0.0166 & 55 & 30.81 \\
		& 0.8 & MLE & 0.0014 & 0.0157 & 100 & 209.97 \\
		&     & Wolter-2 & 0.0007 & 0.0166 & 29.5 & 14.95 \\
		\multicolumn{7}{|c|}{}\\ 
		P3 & 0.4 & MLE & 0.0013 & 0.0103 & 100 & 140.26 \\
		&     & Wolter-2 & 0.0005 & 0.0133 & 53.5 & 24.45 \\
		& 0.8 & MLE & 0.0012 & 0.0090 & 100 & 157.11 \\
		&     & Wolter-2 & 0.0005 & 0.0133 & 21 & 10.59 \\
		\multicolumn{7}{|c|}{}\\ 
		P4 & 0.4 & MLE & 0.0013 & 0.0074 & 100 & 140.14 \\
		&     & Wolter-2 & 0.0008 & 0.0099 & 55.5 & 17.89 \\
		& 0.8 & MLE & 0.0008 & 0.0064 & 100 & 151.99 \\
		&     & Wolter-2 & 0.0008 & 0.0099 & 22 & 5.88 \\
		\multicolumn{7}{|c|}{}\\ 
		P5 & 0.4 & MLE & -0.0001 & 0.0101 & 100 & 161.52 \\
		&     & Wolter-2 & 0.0001 & 0.0184 & 52 & 33.49 \\
		& 0.8 & MLE & 0.0007 & 0.0167 & 100 & 248.26 \\
		&     & Wolter-2 & 0.0001 & 0.0184 & 27 & 17.07 \\
		\multicolumn{7}{|c|}{}\\ 
		P6 & 0.4 & MLE & 0.0013 & 0.0116 & 100 & 361.61 \\
		&     & Wolter-2 & -0.0005 & 0.0248 & 54 & 45.24 \\
		& 0.8 & MLE & 0.0006 & 0.0115 & 100 & 260.15 \\
		&     & Wolter-2 & -0.0005 & 0.0248 & 31 & 25.62 \\
		\hline
	\end{tabular}
	\label{tab:4a}
\end{table}

\begin{table}[ht]
	\scriptsize
	\centering
	\caption{Summary results on the estimators of $N_A$ \& $N_B$ under the simulated model Model II with $(N_A, N_B)=(1200,1000)$ and the ratio of the sub-population sizes ($r=1.2$) is known.}
	\begin{tabular}{|cclrccc|}
		\hline
		\multicolumn{7}{|c|}{}\\
		Population & $\alpha_0$ & \multicolumn{1}{c}{Method} & \multicolumn{1}{c}{RB} & RRMSE & CP($\%$)& LCI\\
		\hline
		\hline
		\multicolumn{7}{|c|}{}\\
		\multicolumn{7}{|c|}{Results on estimators of $N_A$}\\ 
		\hline
		P1 & 0.4 & MLE  & 0.0003 & 0.0061 & 97 & 5.93 \\
		&     & Wolter-1 & -0.0680 & 0.3516  & 53.5 & 1379.14 \\  
		& 0.8 & MLE & 0.0001 & 0.0003 & 90 & 1.80  \\
		&     & Wolter-1  & -0.2125 & 0.6351  & 25.5 & 465.28 \\  
		\multicolumn{7}{|c|}{}\\ 
		P2 & 0.4 & MLE  & $< 10^{-4}$ & 0.0003 & 96.5 & 4.51  \\
		&     & Wolter-1 & -0.0216 & 0.5508 & 71.5 & 1290.38 \\  
		& 0.8 & MLE & $<10^{-4}$ & 0.0015 & 95 & 3.78 \\
		&     & Wolter-1  & -0.0881 & 2.2801  & 34 & 580.83 \\
		\multicolumn{7}{|c|}{}\\ 
		P3 & 0.4 & MLE  & 0.0002 & 0.0033 & 98 & 12.05  \\
		&     & Wolter-1 & 0.2069 & 3.4894  & 81.5 & 923.23 \\  
		& 0.8 & MLE & 0.0004 & 0.0052 & 98 & 15.87  \\
		&     & Wolter-1  & -0.1149 & 0.1825 & 54 & 397.55 \\
		\multicolumn{7}{|c|}{}\\ 
		P4 & 0.4 & MLE  & 0.0005 & 0.0068 & 99 & 14.82  \\
		&     & Wolter-1 & -0.0103 & 0.3202  & 83 & 927.42 \\  
		& 0.8 & MLE & $<10^{-4}$ & 0.0003 & 97.5 & 17.12 \\
		&     & Wolter-1 & -0.0774 & 0.2726  & 52.5 & 420.36 \\
		\multicolumn{7}{|c|}{}\\ 
		P5 & 0.4 & MLE  & 0.0002 & 0.0032 & 96.5 & 1.61  \\
		&     & Wolter-1 & -0.1223 & 0.3675 & 53 & 1643.76 \\  
		& 0.8 & MLE & $<10^{-4}$ & $<10^{-4}$ & 99.5 & 1.28 \\
		&     & Wolter-1  & -0.2561 & 0.8123  & 29.5 & 508.70 \\
		\multicolumn{7}{|c|}{}\\ 
		P6 & 0.4 & MLE  & $<10^{-4}$ & 0.0003 & 99 & 0.91 \\
		&     & Wolter-1 & -0.1910 & 0.4858  & 69 & 1431.90 \\  
		& 0.8 & MLE & $<10^{-4}$ & $<10^{-4}$ & 100 & 3.80  \\
		&     & Wolter-1  & -0.2666 & 0.7618 & 27.5 & 612.75 \\  
		\hline
		\multicolumn{7}{|c|}{}\\
		\multicolumn{7}{|c|}{Results on estimators of $N_B$}\\ 
		\hline
		P1 & 0.4 & MLE  & 0.0002 & 0.0061 & 81.5 & 4.78  \\
		&     & Wolter-1 & -0.0688 & 0.3521  & 54.5 & 1379.15 \\  
		& 0.8 & MLE & -0.0001 & 0.0003 & 89.5 & 1.56  \\
		&     & Wolter-1  & -0.2163 & 0.6371  & 28 & 465.28 \\  
		\multicolumn{7}{|c|}{}\\ 
		P2 & 0.4 & MLE  & $>-10^{-4}$ & 0.0004  & 95.5 & 3.79  \\
		&     & Wolter-1 & -0.0247 & 0.5520   & 67 & 1290.38 \\  
		& 0.8 & MLE & $<10^{-4}$ & 0.0015 & 95 & 3.25 \\
		&     & Wolter-1  & -0.0918 & 2.2806  & 31 & 580.83 \\
		\multicolumn{7}{|c|}{}\\ 
		P3 & 0.4 & MLE  & 0.0002  & 0.0033 & 98 & 10.05 \\
		&     & Wolter-1 & 0.2027 & 3.4896   & 82.5 & 923.23 \\  
		& 0.8 & MLE & 0.0004 & 0.0052 & 98 & 13.24 \\
		&     & Wolter-1  & -0.1194 & 0.1868   & 52 & 397.55 \\
		\multicolumn{7}{|c|}{}\\ 
		P4 & 0.4 & MLE  & 0.0005 & 0.0067  & 99 & 12.33 \\
		&     & Wolter-1 &  -0.0127 & 0.3211  & 85 & 927.42 \\  
		& 0.8 & MLE &  $<10^{-4}$ & 0.0002 & 98.5 & 14.25 \\
		&     & Wolter-1  & -0.0816 & 0.2753 & 51.5 & 420.36 \\
		\multicolumn{7}{|c|}{}\\ 
		P5 & 0.4 & MLE  &  0.0002 & 0.0032  & 97 & 1.38 \\
		&     & Wolter-1 &  -0.1243 & 0.3689   & 52 & 1643.76 \\  
		& 0.8 & MLE & $<10^{-4}$ & $<10^{-4}$ & 99.5 & 1.12 \\
		&     & Wolter-1  & -0.2611 & 0.8149  & 27 & 508.70 \\
		\multicolumn{7}{|c|}{}\\ 
		P6 & 0.4 & MLE  & $<10^{-4}$ & 0.0002 & 99 & 0.78 \\
		&     & Wolter-1 & -0.1982 & 0.4904 & 63.5 & 1431.90 \\  
		& 0.8 & MLE &  $<10^{-4}$ & $<10^{-4}$  & 100 & 3.20 \\
		&     & Wolter-1  & -0.2722 & 0.7651   & 26 & 612.75 \\        
		\hline
	\end{tabular}
	\label{tab:4b}
\end{table}

\section{Applications}\label{realdata}
In this section, we first analyze a data set on Encephalitis (infectious and noninfectious) incidence in England during November 2006 to October 2007 \citep{Granerod13}, presented in the top panel of Table \ref{tab:4}. This particular data was collected adhering to an encephalitis code in any of the 20 diagnostic fields, and segregated into two strata, Children ($<18$ years) and Adult ($\geq 18$ years). A patient detected with encephalitis by a hospital clinician was likely to be recorded in Hospital Episode Statistics (HES) and also included in the the Public Health England (PHE) study. Thus, \citet[p. 1461]{Granerod13} anticipated that the two sources are likely to be positively dependent. As a result, the LP estimator, given in (\ref{LP}), probably underestimate the true number of cases. Note that the estimator proposed by \citet{Nour82} cannot be applied for both the strata as its underlying condition ($x_{11}^2>x_{10}x_{01}$) is not valid. Also, the estimators proposed by \citet{Wolter90} can not be applied as the ratio of adult and child patients (equivalent to sex-ratio for male-female stratification) is not available here. Therefore, we compare the results from our proposed models with that of the LP estimator defined in (\ref{LP}).

As remarked in Subsection \ref{Model01}, the MMEs are approximately equal to the MLEs under Model I, and hence we only consider MLE for our data analysis. For analyzing the data under Model I, we further consider both the cases separately where capture recapture status for Children and Adult are independent. In order to compute the estimate of standard error $\hat{\sigma}_{\hat{N}}$, we use the same parametric bootstrap method as mentioned in Section \ref{simu}. Comparing the relative standard error (RSE), i.e. $\hat{\sigma}_{\hat{N}}\left/\right.\hat{N}$, we find that our proposed estimator under Model I performs better with independent assumption for Children than that for Adult and the corresponding results are reported in the top panel of Table \ref{tab:5}. Estimate of the dependence parameter indicates that $5\%$ of the Adult encephalitis patients are causally dependent. Under Model II, the estimated number of patients is larger compared to that of under Model I. The estimated proportion of causally dependent patients for both Adult and Children are $3\%$ under Model II. It is interesting to note that the relative standard error (RSE), based on 1000 bootstrap samples, of the MLE under Model II (Model I) is substantially smaller compared to those of the MLE under Model I ($M_{t}$ Model).

Now we consider another dual system dataset (\textit{See} bottom panel of Table \ref{tab:4}) from Wagai and Yala divisions in western Kenya on child mortality, named as \textit{Gem} in the article by \citet{Eisele03}. This study is on the completeness and differential ascertainment of vital events related to child health among male and female children (less than five years old) registered in demographic surveillance system (DSS) based on two-sample capture-recapture experiment. Here also, both the methods, proposed by \citet{Wolter90} and \citet{Nour82}, are not applicable because of same reasons mentioned earlier. Analyzing the data, we find that performance of our proposed estimator $\hat{N}_{k}$s under Model I, for $k=\text{Male, Female}$, performs better with the assumption that capture recapture status for Female are independent. The results are presented in the bottom panel of Table \ref{tab:5}. It is seen that the estimates for female deaths based on Model I and $M_t$ model are very close, however the r.s.e is for Model I is smaller compared to that of $M_t$ model. Estimate of the dependence parameter indicates that $7\%$ of male child are causally dependent under Model I. Under Model II, the MLEs are marginally lower compared to those of under Model I. In this case the RSE of the estimates under Model I(Model II) is smaller compared to those under Model II ($M_{t}$ model). Based on our analysis no evidence of list-dependence was found in the DRS under consideration which supports the argument made by \citet{Eisele03}.       

\begin{table}[ht]
	\centering
	\caption{Data sets used in illustration of the proposed methods.}
	\resizebox{3.5in}{!}{
		\begin{tabular}{|llcccc|}
			\hline
			Dataset&Stratum& $x_{11}$ & $x_{10}$ & $x_{01}$ & Total\\
			\hline
			\hline
			\multicolumn{6}{|c|}{}\\
			Encephalitis & Adult & 39 & 290 & 39& 368\\
			& Children  & 20 & 78 & 15 & 113\\
			\multicolumn{6}{|c|}{}\\
			Children Death & Male &30 &153& 8 & 191\\
			&Female& 15 & 173 & 7 & 195\\
			\hline
		\end{tabular}
	}
	\label{tab:4}
\end{table}    

\begin{table}[!ht]
	\centering
	\caption{Summary results of real data analysis with proposed Model I and II.}
	\begin{minipage}{12cm}
		\begin{tabular}{|cccccc|}
			\hline
			& & & Model I & Model II & LP \\ 
			Dataset & Stratum & & MLE & MLE & $\left(\hat{N}^{(LP)}\right)$ \\
			\hline\hline
			& Adult & $\hat{N}_{A}$[RSE]  & 660 [0.077] & 739 [ 0.012] & 658 [0.212] \\
			&& C.I.   & (563,  760)  & (731, 748) & (463, 988) \\
			Encephalitis & & $\hat{\alpha}_{A}$ & 0.052 & 0.031 & -  \\
			&Children & $\hat{N}_{B}$[RSE]& 197 [ 0.104] & 213[0.072] & 171[ 0.317] \\
			&& C.I.  & (160, 241)  & (160, 241) & (101, 314) \\
			\hline
			&&   &   &  &  \\
			&Male & $\hat{N}_{A}$[RSE] & 268 [0.054] & 250 [0.092] & 231[0.244]  \\
			&& C.I.  & (244, 303)  & (204, 302) & (151, 362) \\
			Children Death & & $\hat{\alpha}_{A}$ & 0.070 & 0.006 & - \\
			&Female & $\hat{N}_{B}$[RSE] & 276 [0.052] & 262 [0.097] & 275 [0.424]  \\
			&& C.I.& (250, 306)  & (212, 324) & (145, 552) \\
			\hline
		\end{tabular} 
	\end{minipage}
	\label{tab:5}
\end{table}

\section{Concluding Remarks}\label{conclusions}
This article deals with a very interesting problem when causal independence assumption in DRS is not valid. We introduce a model, called Bivariate Bernoulli model, that successfully accounts for the possible dependence between capture and recapture attempts. Though the proposed model discusses positive correlation, one can rewrite the model easily in order to incorporate negative dependence (See Ramark \ref{mark_1}). Our proposed model seems to have an edge in terms of ease of interpretation and has much wider domain of applicability. In case, the ratio of the subpopulation sizes (e.g., sex-ratio for male-female stratification) is known, estimates based on our proposed models may be preferred. This also allows inclusion of any additional information (e.g. sex-ratio), if available, to make more efficient inference. Although the primary objective of this article is to obtain an efficient estimate of the population size $N$, the estimates of the other model parameters, especially $\hat{\alpha}$, give specific insights into the capture-recapture mechanism. The BBM can also be extended for multiple list or multiple capture-recapture problems which is commonly encountered in the study of wildlife population. It is also an interesting problem to develop a testing procedure to test the behavioral dependence between two sources in DRS, which will be taken up in future work.

\section*{Appendix}

\textbf{Derivation for MME under Model I:}\\
We get from (\ref{MME_MI}), the following equation in terms of $p_{2A}$, $\alpha_{A}$, and $N_{A}$:
\begin{eqnarray}
N_{A}\alpha_{A}\hat{p}_{1} +\left(1-\alpha_{A}\right)N_{A}\hat{p}_{1}p_{2A} & = & x_{11A},\label{c1}\\
N_{A}\hat{p}_{1}\left(1-p_{2A}\right)\left(1-\alpha_{A}\right) & = & x_{10A}\label{c2}\\
N_{A}p_{2A}\left(1-\hat{p}_{1}\right)\left(1-\alpha_{A}\right) & = & x_{01A}\label{c3},
\end{eqnarray}
where $\hat{p}_1=\hat{p}_{1\cdot A}=\frac{x_{11B}}{x_{\cdot1B}}$.
Now, by adding (\ref{c1}) and (\ref{c2}), we get the MME of $N_{A}$ as
\begin{eqnarray}
\hat{N}_{A} & = & \frac{x_{1\cdot A}x_{\cdot1B}}{x_{11B}}.\nonumber
\end{eqnarray}

Again, by adding the equations (\ref{c1})-(\ref{c3}), \begin{eqnarray}
N_{A}\hat{p}_1+N_{A}p_{2A}\left(1-\alpha_{A}\right)\left(1-\hat{p}_1\right)&=&x_{0A}\label{c5}
\end{eqnarray}
and by subtracting (\ref{c3}) from (\ref{c2}), we get
\begin{eqnarray}
N_{A}\left(1-\alpha_{A}\right)\left(\hat{p}_1-p_{2A}\right)&=&\left(x_{10A}-x_{01A}\right).\nonumber\label{eqn_subtract}
\end{eqnarray}

Now, using the estimates $\hat{N}_{A}$ and $\hat{p}_1$ in (\ref{c5}), we get
\begin{eqnarray}
p_{2A}\left(1-\alpha_{A}\right)&=&\frac{x_{01A}x_{11B}}{x_{01B}x_{1\cdot A}}.\label{c6}
\end{eqnarray}

Since $N_{A}\hat{p}_{1}=x_{1\cdot A}$, (\ref{c1}) implies
\begin{eqnarray}
\alpha_{A}+p_{2A}\left(1-\alpha_{A}\right)&=&\frac{x_{11A}}{x_{1\cdot A}}.\label{c7}
\end{eqnarray}

Subtracting (\ref{c6}) from (\ref{c7}), the MME of $\alpha_{A}$ is obtained as
\begin{eqnarray}
\hat{\alpha}_{A} & = & \frac{x_{\cdot1A}}{x_{1\cdot A}}-\frac{x_{01A}x_{\cdot1B}}{x_{01B}x_{1\cdot A}}.\label{c8}
\end{eqnarray}
Using $\hat{\alpha}$ in (\ref{c6}), MME of $p_{2A}$ is given as
\begin{eqnarray}
\hat{p}_{2A} & = & \frac{x_{01A}x_{11B}}{x_{10A}x_{01B}+x_{01A}x_{11B}}.\nonumber
\end{eqnarray}

In order to ensure that MME of $\alpha_{A}$ lies in $[0,1]$, we modify (\ref{c8}) and consider
$$\hat{\alpha}_{A}=\min\left\{\max\left\{0,\frac{x_{\cdot1A}}{x_{1\cdot A}}-\frac{x_{01A}x_{\cdot1B}}{x_{01B}x_{1\cdot A}}\right\},1\right\}.$$
\hspace{5.9in}$\Box$\\
\\

\textbf{Derivation for MME under Model II:}
We get from (\ref{MME_MI}), the following equation in terms of $p_{1A}$, $p_{2A}$, $\alpha_{A}$, and $N_{A}$:
\begin{eqnarray}
N_{A}p_{1A}\left(1-p_{2A}\right)\left(1-\alpha_{A}\right) & = & x_{10A},\label{c22}\\
N_{A}p_{2A}\left(1-p_{1A}\right)\left(1-\alpha_{A}\right) & = & x_{01A}\label{c32}.
\end{eqnarray}
Now, dividing (\ref{c22}) by (\ref{c32}) we get
\begin{eqnarray}
\frac{p_{1A}\left(1-p_{2A}\right)}{p_{2A}\left(1-p_{1A}\right)}& = & \frac{x_{10A}}{x_{01A}}\label{c42}.
\end{eqnarray}
Next, we equate the expected and observed number of cell counts from the 2$\times$2 table obtained under DRS for the sub-population $U_B$ and get
\begin{eqnarray}
N_{B}p_{1B}\left(1-p_{2B}\right)\left(1-\alpha_{B}\right) & = & x_{10B},\label{c62}\\
N_{B}p_{2B}\left(1-p_{1B}\right)\left(1-\alpha_{B}\right) & = & x_{01B}.\label{c72}
\end{eqnarray}
Now, we consider the assumption (\textit{S4}) and $\alpha_A=\alpha_B=\alpha_0$. Therefore, dividing (\ref{c22}) by (\ref{c62}) we get
\begin{eqnarray}
\frac{N_{A}\left(1-p_{2A}\right)}{N_{B}\left(1-p_{2B}\right)}& = & \frac{x_{10A}}{x_{10B}}\nonumber\\
\Rightarrow\frac{1-p_{2A}}{1-p_{2B}}& = & \frac{x_{10A}}{x_{10B}}\frac{x_{1\cdot B}}{x_{1\cdot A}}\label{c82},
\end{eqnarray}		
since $N_A=\frac{x_{1\cdot A}}{x_{1\cdot B}}N_B$.\\ 
Similarly, dividing (\ref{c32}) by (\ref{c72}) we get
\begin{eqnarray}
\frac{N_{A}p_{2A}\left(1-p_{1A}\right)}{N_{B}p_{2B}\left(1-p_{1B}\right)}& = &\frac{x_{01A}}{x_{01B}}\nonumber\\
\Rightarrow\frac{p_{2A}}{p_{2B}}& = & \frac{x_{01A}}{x_{01B}}\frac{x_{1\cdot B}}{x_{1\cdot A}}\nonumber\\
\Rightarrow p_{2A}& = & \frac{x_{01A}}{x_{01B}}\frac{x_{1\cdot B}}{x_{1\cdot A}}p_{2B}\label{c92}.
\end{eqnarray}
From equations (\ref{c82}) and (\ref{c92}), we get
\begin{eqnarray}
1-\frac{x_{10A}x_{1\cdot B}}{x_{10B}x_{1\cdot A}}& = &   p_{2B}\left(\frac{x_{01A}x_{1\cdot B}}{x_{01B}x_{1\cdot A}}-\frac{x_{10A}x_{1\cdot B}}{x_{10B}x_{1\cdot A}}\right)\nonumber\\
\Rightarrow\hat{p}_{2B}& = & \frac{x_{01B}}{x_{1\cdot B}}\frac{x_{10B}x_{1\cdot A}-x_{10A}x_{1\cdot B}}{x_{01A}x_{10B}-x_{10A}x_{01B}}\nonumber
\end{eqnarray}
and
\begin{eqnarray}
\hat{p}_{2A}& = & \frac{x_{01A}}{x_{1\cdot A}}\frac{x_{10B}x_{1\cdot A}-x_{10A}x_{1\cdot B}}{x_{01A}x_{10B}-x_{10A}x_{01B}}\nonumber.
\end{eqnarray}
Therefore, by putting the above estimate $\hat{p}_{2A}$ in equations (\ref{c22}) and (\ref{c42}), we get
\begin{eqnarray}
\hat{\alpha}& = & 1-\frac{x_{10A}}{x_{1\cdot A}}\frac{1}{1-\hat{p}_{2A}}\nonumber
\end{eqnarray}
and 
\begin{eqnarray}
\hat{p}_{1A}& = & \frac{1}{1+\frac{x_{01A}}{x_{10A}}\left(\frac{1}{\hat{p}_{2A}}-1\right)},\nonumber
\end{eqnarray}
respectively. Finally, we obtain the estimates of sub-populations sizes as
$$\hat{N}_{A}=\frac{x_{1A}}{\hat{p}_{1A}},  \hat{N}_{B}=\frac{x_{1B}}{\hat{p}_{1B}},$$
since $N_{A}p_{1A}=x_{1\cdot A}$ and $N_{B}p_{1B}=x_{1\cdot B}$.

\bibliographystyle{apalike}
\bibliography{sample}

\end{document}